\documentclass[letterpaper, 10 pt, conference]{ieeeconf}  

\IEEEoverridecommandlockouts                              

\usepackage{graphics} 
\usepackage{epsfig} 
\usepackage{amsmath} 
\usepackage{amssymb}  
\usepackage{cite}
\usepackage{float}
\usepackage{mathrsfs}
\usepackage{mathtools}
\usepackage{svg}
\usepackage{times} 
\usepackage{subcaption}
\usepackage{comment}
\newtheorem{theorem}{Theorem}[section]

\newtheorem{lemma}[theorem]{Lemma}
\newtheorem{corollary}[theorem]{Corollary}

\newtheorem{definition}{Definition}

\newcommand{\leftbracket}{\left[}

\newcommand{\rightbracket}{\right]}

\title{\Large \bf
Altruism Improves Congestion in Series-Parallel Nonatomic Congestion Games
}

\author{Colton Hill and Philip N. Brown
\thanks{*This material is based upon work supported by the National Science Foundation under award number ECCS-2013779.}
\thanks{The authors are with the University of Colorado at Colorado Springs, CO 80918, USA.
	{\tt\small \{chill13, pbrown2\}@uccs.edu}}%
}

\begin{document}

\maketitle
\thispagestyle{empty}
\pagestyle{empty}

\begin{abstract}

Self-interested routing polices from individual users in a system can collectively lead to poor aggregate congestion in routing networks.
The introduction of altruistic agents, whose goal is to benefit other agents in the system, can seemingly improve aggregate congestion.
However, it is known in that in some network routing problems, altruistic agents can actually worsen congestion compared to that which would arise in the presence of a homogeneously selfish population.
This paper provides a thorough investigation into the necessary conditions for altruists to be guaranteed to improve total congestion.
In particular, we study the class of series-parallel non-atomic congestion games, where one sub-population is altruistic and the other is selfish.
We find that a game is guaranteed to have improved congestion in the presence of altruistic agents (even if only a small part of the total population) compared to the homogeneously selfish version of the game, provided the network is \textit{symmetric}, where all agents are given access to all paths in the network, and the series-parallel network for the game does not have sub-networks which emulate Braess's paradox --- a phenomenon we refer to as a \textit{Braess-resistant} network.
Our results appear to be the most complete characterization of when behavior that is designed to improve total congestion (which we refer to as altruism) is actually guaranteed to do so.

\end{abstract}

\section{INTRODUCTION}\label{sec:Intro}

As society and technology become increasingly connected, there is an increasing requirement to understand and improve the coordination between human social behavior and the technological operations within these systems~\cite{ratliffPerspectiveIncentiveDesign2019}.
The increasing prevalence of autonomous vehicles has presented an another vessel in which system designers can optimize traffic routing in transportation networks, and is thus a canonical (and significant) example of this interaction.
It is widely recognized that when individuals choose routes solely to minimize their own travel time, it can lead to suboptimal congestion~\cite{roughgardenSelfishRoutingPrice2005}. 
Game theoretic concepts provide systematic methods that have been widely studied to compare the inefficiencies resulting from the behaviors of self-interested agents with that of the optimal aggregate congestion for a network~\cite{roughgardenIntrinsicRobustnessPrice2015}.

Centralized routing control is one method to reduce network inefficiencies, although it is often highly infeasible, so many studies have aimed to improve network efficiency in a less-centralized manner.
Techniques designed to influence behavior have been studied such as direct fleet routing strategies~\cite{roughgardenStackelbergSchedulingStrategies2001}, information distribution~\cite{chengInformationalNudgingBoundedly2018}, and positive or negative monetary incentives, such as tolls and subsidies~\cite{fergusonCarrotsSticksEffectiveness2020}.
Additionally, a system designer for a fleet of autonomous vehicles may design the fleet's routing policies to be altruistic, so they consider their own contribution to congestion~\cite{brownWhenAltruismWorse2021, hill2023tradeoff}.
These concepts are akin to asking when \textit{altruistic} agents, who consider their own contribution to aggregate congestion, improve total latency compared to agents who only consider minimizing their own travel time (\textit{selfish} agents).

An intuitive way to measure the inefficiency that arises in the presence of an all-selfish population is to take the ratio of their total latency with that of the optimal traffic latency.
This idea is formalized by the concept of \textit{price of anarchy}, and is defined as the ratio between the total latency in a system where all agents act selfishly and choose roads that minimizes their commute time (modeled by a \emph{Nash equilibrium}) with the total latency that can be achieved if a system designer centrally coordinates all agents' routing for the overall benefit of the system~\cite{roughgardenIntrinsicRobustnessPrice2015}.
If the price of anarchy is $1$, traffic is routing optimally, and as it becomes greater than $1$, traffic is known to be routing sub-optimally.
The price of anarchy serves as an upper bound on how much a system designer could potentially improve aggregate congestion if she is able to centrally route all traffic.

A promising way to address inefficiency regarding selfish traffic is from the perspective of a system designer choosing the routing policy for a fleet of autonomous vehicles. If the designer chooses an \emph{altruistic} routing policy, where all vehicles choose routes in consideration of their impact on aggregate road congestion, it is known that all inefficiency would be eliminated~\cite{chenAltruismItsImpact2014, liEmployingAltruisticVehicles2021}.
However, it has been shown that if the fleet the system designer is responsible for only makes up a fraction of the total population, the total latency that arises from the heterogeneous population can actually be worse than if all traffic had remained selfish~\cite{brownCanTaxesImprove2020}.
Again, we can precisely measure the inefficiency that may arise from altruism by taking the ratio of the total latency that arises when all agents route to minimize the latency they experience as Nash equilibrium, with the total latency that arises when all agents route selfishly; a ratio known as the \textit{perversity index}~\cite{brownCanTaxesImprove2020}.
In contrast to the price of anarchy, the perversity index can be less than $1$, in which case altruists improve congestion relative to homogeneous selfishness.
But, as the perversity index becomes greater than $1$, heterogeneous altruism actually degrades welfare compared to homogeneous selfishness.
That is, partially-adopted altruism has the potential to actually degrade congestion relative to an all-selfish population~\cite{brownWhenAltruismWorse2021}.

Thus, the key motivation for our work is to characterize the topological network requirements so that altruistic agents are guaranteed not to degrade overall congestion.
That is, what constraints must be put on the network so that the perversity index is less than $1$ for a heterogeneous population of altruistic and selfish agents.
Previous work has provided extensive results on what network topology produces efficient equilibria~\cite{milchtaichNetworkTopologyEfficiency2006}, and has shown that altruism cannot degrade congestion on serially-linearly-independent networks with affine cost functions~\cite{sekarUncertaintyMulticommodityRouting2019}.

A summary of our results is as follows:
\begin{enumerate}
    \item Theorem~\ref{thm:Main} shows that in a game with a heterogeneous population of altruistic and selfish agents, if the network is symmetric (all agents have access to all paths), and Braess-resistant, then an increase in the fraction of altruistic agents in the network is guaranteed to improve the total latency of the network, provided the network is series-parallel.
   
    \item Lemma~\ref{lem:Milchtaich_Extension} provides an extension of~\cite[Lemma 3.3]{milchtaichNetworkTopologyEfficiency2006}, and Corollary~\ref{cor:p_a_AND_p_s} applies the lemma specifically to our context.
    They show that the experienced path cost functions for agents of either type is monotone as a function of their sub-population size.
    Specifically, if one sub-population increases, and the other sub-population decreases by the same amount, the path cost increases for agents of the increasing sub-population, and the the path cost decreases for agents of the decreasing sub-population.

    \item Lemma~\ref{lem:Path_Ordering} shows that path flows are monotone with respect to marginal changes in the fraction of populations.
    This shows that even small decreases in altruistic population size (and subsequent increases in selfish population) can increase path flows for other selfish agents.
\end{enumerate}

\noindent\textit{Organization:} The paper is organized as follows.
Section~\ref{sec:Model} provides the nonatomic congestion game model in the presence of heterogeneous types and provides related work.
Section~\ref{sec:Contribution} provides the main result, and Section~\ref{sec:Proof} provides the proof of the main result, along with the statement of each supporting result.
Section~\ref{sec:CounterExamples} provides negative examples for networks that do not satisfy the necessary symmetric and Braess-resistant conditions.
Section~\ref{sec:Conclusion} concludes.

\section{Model and Related Work}\label{sec:Model}

\subsection{Routing Problem}

We consider a routing problem for a network $(V, E)$, consisting of vertex set $V$ and edge set $E$.
Each edge $e \in E$ connects two distinct vertices, and we say that an edge $e$ is \textit{incident} with vertex $v \in V$ if $v$ is an end vertex of $e$.
A sub-path $\sigma$ of length $n \geq 0$ is an alternating sequence of distinct vertices and $n$ edges, beginning and ending with vertices, where each edge $e \in \sigma$ is incident with the vertex preceding and the vertex succeeding $e$.
Also, we say the vertices that are incident with $\sigma$ are the first and last vertex of $\sigma$.
A path $p$ comprises a set of edges connecting common origin $o$ to common destination $t$, and can also be represented by a set of sub-paths, where each sub-path $\sigma \in p$ is a series of edges, with each edge $e \in p$.
We write $\mathcal{P} \subseteq 2^E$ to denote the set of \textit{paths} accessible to agents, and $\mathcal{R} \subseteq 2^E$ to denote the set of \textit{all} feasible paths connecting $o$ to $t$, and is referred to as the set of \textit{routes}.
We restrict network topology to \textit{series-parallel}; a network is series-parallel if it is 
\begin{enumerate}
    \item a single edge, 
    \item two series-parallel networks connected in series, or
    \item two series-parallel networks connected in parallel~\cite{milchtaichNetworkTopologyEfficiency2006}.
\end{enumerate}

A unit mass of traffic is routed from $o$ to $t$ and is composed of two types, those belonging to an \textit{altruistic} population and the remaining uninfluenced traffic is referred to as \textit{selfish}.
Altruistic agents comprise mass $r^{\rm{a}}$, and selfish users make up mass $r^{\rm{s}}$, such that $r^{\rm{a}} + r^{\rm{s}} = 1$.
Each population of type $\theta \in \{\rm{a}, \rm{s}\}$ traffic can access an arbitrary subset of paths $\mathcal{P}^\theta \subseteq \mathcal{P}$, where $x^\theta_p$ denotes the flow of agents of type $\theta$ using path $p \in \mathcal{P}^\theta$.
A \textit{feasible flow} for type $\theta$ is an assignment of $r^\theta$ mass of traffic to paths in $\mathcal{P}^\theta$, denoted by $x^\theta \in \mathbb{R}^{|\mathcal{P}^\theta|}_{\geq 0}$, such that $\sum_{p \in \mathcal{P}} x^\theta_p = r^\theta$.
A \textit{network flow} is a combined allocation of altruistic and selfish agents to paths, denoted $x \in \mathbb{R}^{|\mathcal{P}|}_{\geq 0}$, such that $x_p \coloneqq \sum_{\theta : p \in \mathcal{P}^\theta}x^\theta_p$, where the flows $x^\theta$ are feasible for their respective types.
Provided a network flow $x$, the flow on edge $e \in E$ is given by $x_e = \sum_{\sigma : e \in \sigma} x_\sigma = \sum_{p : e \in p} x_p$, and we denote the flow of type $\theta = \{\rm{a}, \rm{s}\}$ on edge $e$ by $x^\theta_e$.
For each edge $e \in E$, commute time is expressed as a function of traffic flow and is associated with a latency function $\ell_e : [0, 1] \rightarrow [0, \infty)$, where $\ell_e(x_e)$ is the cost experienced by agents on edge $e$ with edge flow $x_e$.
We assume the latency function for each edge is a non-decreasing, convex posynomial (i.e. a polynomial with non-negative coefficients).
So, for every $e \in E$,
\begin{equation}\label{eq:LatencySumDef}
    \ell_e(x_e) = \sum_{i=0}^d a_{e, i}x^i_e,
\end{equation}
where $d \in \mathbb{N}$, and $a_{e,i} \in \mathbb{R}_{\geq 0}$.

We measure the cost of a flow by the \textit{total latency}, given by
\begin{equation}\label{eq:SocialCost}
    \mathcal{L}(x) = \sum_{e \in E} x_e\ell_e(x_e) = \sum_{p \in \mathcal{P}} x_p\ell_p(x),
\end{equation}
where we define the latency of path $p$, given flow $x$, as the sum of the latencies of each edge $e \in p$:
\begin{equation}\label{eq:pathLatency}
    \ell_p(x) \coloneqq \sum_{e \in p} \ell_e(x_e).
\end{equation}
Similarly, we define the latency of a sub-path $\sigma$, given flow $x$, as the sum of latencies of each edge in $e \in \sigma$:
\begin{equation}\label{eq:sub-pathLatency}
    \ell_\sigma(x) \coloneqq \sum_{e \in \sigma} \ell_e(x_e).
\end{equation}

An instance of a \textit{routing problem} is fully specified by the tuple $G = (V, E, \{\ell_e\}, \mathcal{P}^{\rm{s}}, \mathcal{P}^{\rm{a}}, r^{\rm{a}})$, and we write $\mathcal{G}(d, r^{\rm{a}})$ to denote the set of all routing problems on series-parallel networks with posynomial latency functions of degree at most $d$ and altruistic population $r^{\rm{a}}$.

\subsection{Heterogeneous Routing Game}

To understand how altruistic agents effect congestion within the context of a heterogeneous population, we model the routing problem as a nonatomic congestion game.
Each type of traffic is composed of infinitely many infinitesimal agents, where the cost experienced by agents is determined by their type.
Given a flow $x$, the cost a selfish agent experiences for using path $p \in \mathcal{P}^{\rm{s}}$ is simply the actual latency of the path:
\begin{align}\label{eq:SelfishPathCost}
    \nonumber
    \ell_p^{\rm{s}}(x) &\coloneqq \sum_{e \in p} \ell_e(x_e)
    \\ &=
    \ell_p(x).
\end{align}
Intuitively, (\ref{eq:SelfishPathCost}) assumes selfish agents are only interesting in reducing their own commute time, and are uniform in regard to their routing policy.

The cost experienced by altruistic agents considers both the commute time of the agent, as well as their contribution to congestion along the path they are using.
That is, the cost an altruistic agent experiences for using path $p \in \mathcal{P}^{\rm{a}}$ is the marginal cost of the path:
\begin{align}\label{eq:MarginalPathCost}
    \nonumber
    \ell^{\rm{a}}_p(x) &\coloneqq \sum_{e \in p} \leftbracket \ell_e(x_e) + x_e \ell^\prime_e(x_e) \rightbracket
    \\ &=
    \ell^{\rm{mc}}_p(x),
\end{align}
where $\ell^\prime$ denotes the flow derivative of $\ell$.

Each agent travels from origin $o$ to destination $t$ using the minimum-cost path from those available in their path set.
We call a flow $x$ a \textit{Nash flow} if all agents are individually using minimum-cost paths relative to the choices of others.
That is, for each type $\theta \in \{\rm{a}, \rm{s}\}$, there exists a feasible $x^\theta$ such that the following condition is satisfied:
\begin{equation}\label{eq:Nashflow}
    \forall p, p^\prime \in \mathcal{P}^\theta, x_p^\theta > 0 \Longrightarrow \ell_p^\theta(x) \leq \ell_{p^\prime}^\theta(x).
\end{equation}
Further, it is well known that a Nash flow for any nonatomic congestion game of the aforementioned structure exists~\cite{mas-colellTheoremSchmeidler1984}.
Moreover, for a given Nash flow $x$, we denote the common path cost for users of each type $\theta \in \{\rm{a}, \rm{s}\}$ as $\Lambda_\theta(x)$.

Given routing problem $G$, we write $\tilde{G}$ to denote the \textit{marginalized} version of $G$.
$\tilde{G}$ is identical to $G$ except \mbox{$\varepsilon \in (0, \arg\min_{p \in \mathcal{P}^{\rm{a}}} \{x_p^{\rm{a}} > 0\})$} mass of altruistic traffic adopts the selfish cost function (i.e. $\tilde{r}^{\rm{a}} = r^{\rm{a}} - \varepsilon$ and $\tilde{r}^{\rm{s}} = r^{\rm{s}} + \varepsilon$).
Path sets, latency functions, and network topology remain the same.

\subsection{Related Work}

The inefficiency of selfish routing in congestion games has been extensively studied~\cite{roughgarden2007routing}.
In particular, there has been a focus on the cost at equilibrium as a function of network topology~\cite{roughgarden2002price}, and the degree of cost functions~\cite{roughgarden2002bad}.
Moreover, it is known that in series-parallel networks with affine cost functions and a homogeneously altruistic population ($r^{\rm{a}} = 1$), the total latency is guaranteed to be better than when the population is homogeneously selfish~\cite{sekarUncertaintyMulticommodityRouting2019}.
Further, it is known that altruism can produce unbounded improvements over selfishness in parallel networks~\cite{biyik2019green}.
In parallel networks with symmetric path sets ($\mathcal{P}^{\rm{a}} = \mathcal{P}^{\rm{s}} = \mathcal{P}$) heterogeneous altruism both lowers the price of anarchy compared to homogeneous selfishness~\cite{chenAltruismItsImpact2014}, and the perversity index is unity in these networks~\cite{brownFundamentalLimitsLocallycomputed2017}.

These results do not extend to series-parallel networks, however, as altruism can cause unbounded harm if the pathsets are not symmetric ($\mathcal{P}^{\rm{a}} \neq \mathcal{P}$)~\cite{brownWhenAltruismWorse2021}.
However, partial altruism can be utilized to mitigate the risk of harm caused by selfishness and heterogeneous altruism~\cite{hill2023tradeoff}.
It is known that in serially-linearly-independent networks, heterogeneous altruism is guaranteed to improve network efficiency~\cite{sekarUncertaintyMulticommodityRouting2019}.
The equilibrium costs of users is also monotone with respect to overall population size (even if the path set is not symmetric), provided all users have the same cost functions~\cite{cominetti2024monotonicity}.

\section{Contribution: Altruism is Guaranteed to Improve Congestion in Braess-Resistant and Symmetric Networks}\label{sec:Contribution}

It is well-known that altruism can cause inefficiency with respect to the total latency of a network.
These inefficiencies often arise because of characteristics regarding network topology and road access to altruistic agents.
Hence, we seek a set of conditions that guarantees the presence of altruists will improve total latency.

Our first definition addresses the fact that if altruistic and selfish agents do not have the same path set, altruism can potentially harm the system welfare.

\begin{definition}\label{def:Symmetric}
    A game $G$ is \textit{symmetric} if 
    \begin{equation}
        \mathcal{P}^{\rm{a}} = \mathcal{P}^{\rm{s}} = \mathcal{P}.
    \end{equation}
\end{definition}

Next, it can be observed that series-parallel networks can have sub-networks which behave like Wheatstone networks, and thus emulate Braess's paradox.
Our next definition clarifies the intention of series-parallel networks, and precludes sub-networks which produce Braess's paradox.
\begin{definition}\label{def:Braess_Resistant}
    A game $G$ is \textit{Braess-resistant} if $\mathcal{P} = \mathcal{R}$.
\end{definition}
Networks that don't satisfy either of these conditions can be shown to have perversity; examples are given in Section~\ref{sec:CounterExamples}.

Our main result shows that any network that is both Braess-resistant and symmetric is guaranteed to have improved total latency in the presence of altruism (or likewise, worsen total latency if existing altruists become selfish).

\begin{theorem}\label{thm:Main}
    Let $x$ and $\tilde{x}$ be a Nash flow for a game $G$ and its marginal counterpart, $\tilde{G}$, respectively.
    If $G$ (and by proxy, $\tilde{G}$) is both Braess-resistant and symmetric, then 
    \begin{equation}
        \mathcal{L}(x) \leq \mathcal{L}(\tilde{x}).
    \end{equation}
\end{theorem}

The proof is completed in Section~\ref{sec:Proof}; we provide intuition for the result, and discuss its consequences here.
The importance of Theorem~\ref{thm:Main} is twofold; it is strikingly simple, and has an extremely wide breadth applications.
The result simply says that, for a game $G$ with altruistic population $r^{\rm{a}}$, if the fraction of altruists is decreased even slightly, the resulting cost of congestion is guaranteed to weakly worsen.
Likewise, since the proof is completed by calculating the subgradient of $\mathcal{L}(x)$, it also implies that even a slight increase in altruistic population as a fraction of total traffic is guaranteed to weakly improve congestion.
Hence, the impact of this result has a wide range of applications.
A system planner for a set of roads can use this result to design the topology of the network so that subsequent tolls, system designers for fleets of autonomous vehicles, and any form of altruistic agent can be certain that their presence improves efficiency of the network.
In the next section, we present the supporting material for Theorem~\ref{thm:Main}, then provide its proof.

\section{Proof of Theorem~\ref{thm:Main}}\label{sec:Proof}
The proof of Theorem~\ref{thm:Main} is completed in four steps:
\begin{enumerate}
    \item Lemma~\ref{lem:Milchtaich_Extension} is an extension of in~\cite[Lemma 3.3]{milchtaichNetworkTopologyEfficiency2006}, and demonstrates that if the total origin-destination flow for two feasible vectors can be ordered, and the total origin-destination flow for the two feasible vectors can be ordered in the same way for a specific population, then there exists a minimum cost path such that each edge in the path has the same ordering as the vectors.
    
    \item Corollary~\ref{cor:p_a_AND_p_s} specifies Lemma~\ref{lem:Milchtaich_Extension} specifically for our context, and concludes that the commonly experienced path cost for altruists weakly decreases from $x$ to $\tilde{x}$, and the commonly experienced path cost for selfish agents weakly increases from $x$ to $\tilde{x}$.
    
    \item Lemma~\ref{lem:Path_Ordering} shows that path flows weakly decrease from $x$ to $\tilde{x}$ for altruists, and weakly increase from $x$ to $\tilde{x}$ for selfish agents.

    \item The proof of Theorem~\ref{thm:Main} provides a lower bound for the subgradient $\frac{\partial}{\partial \varepsilon} \mathcal{L}(x(\varepsilon))$, where $\varepsilon$ is the decrease in altruistic population from $G$ to $\tilde{G}$. 
\end{enumerate}
We state the supporting material here, and proceed with their proofs in the Appendix.

\begin{lemma}\label{lem:Milchtaich_Extension}
    Fix $\theta \in \{\rm{a}, \rm{s}\}$, and let $\hat{f}$ and $\tilde{f}$ be feasible flows for $G$.
    If the path set $\mathcal{P}$ is Braess-resistant and symmetric, \mbox{$||\hat{f}||_1 \geq ||\tilde{f}||_1$}, and $||\hat{f}^\theta||_1 \geq ||\tilde{f}^\theta||_1$ such that $||\hat{f}^\theta|| > 0$, then there is some path $p$ such that for each edge $e \in p$, $\hat{f}_e^\theta \geq \tilde{f}_e^\theta$ and $\hat{f}_e^\theta > 0$.
\end{lemma}
Intuitively, Lemma~\ref{lem:Milchtaich_Extension} says that if an increase in either the altruistic population or selfish population corresponds to an increase in total population, then there is a path used (by the same population) in the increased flow such the cost of the path also increases.
We make the results of Lemma~\ref{lem:Milchtaich_Extension} explicit with the following corollary.

\begin{corollary}\label{cor:p_a_AND_p_s}
    Let $x$ and $\tilde{x}$ be a Nash flow for $G$ and $\tilde{G}$, respectively.
    Since \mbox{$||x||_1 = ||\tilde{x}||_1$}, $||x^{\rm{a}}||_1 \geq ||\tilde{x}^{\rm{a}}||_1$, and $||x^{\rm{s}}||_1 \leq ||\tilde{x}^{\rm{s}}||_1$, there exist paths (which we denote $p_{\rm{a}}$ and $p_{\rm{s}}$), such that for each edge $e \in p_{\rm{a}}$
    \begin{equation}\label{eq:p_a}
        x_e^{{\rm{a}}} \geq \tilde{x}_e^{\rm{a}} \text{ and } x_e^{{\rm{a}}} > 0,
    \end{equation}
    and for each edge $e \in p_{\rm{s}}$
    \begin{equation}\label{eq:p_s}
        x_e^{{\rm{s}}} \leq \tilde{x}_e^{\rm{s}} \text{ and } \tilde{x}_e^{{\rm{s}}} > 0.
    \end{equation}
    Furthermore, $\Lambda_{\rm{a}}(x) \geq \Lambda_{\rm{a}}(\tilde{x})$ and $\Lambda_{\rm{s}}(x) \leq \Lambda_{\rm{s}}(\tilde{x})$.
\end{corollary}

Intuitively, it follows that the total origin-destination flow for $x$ and $\tilde{x}$ is equal.
Thus, since $r^{\rm{a}} > r^{\rm{a}} - \varepsilon$, there exists a path in $x$ used by altruists use such that each edge along that path is non-increasing in flow in $\tilde{x}$, which we denote $p_{\rm{a}}$.
Similarly, since $r^{\rm{s}} < r^{\rm{s}} + \varepsilon$, there exists a path in $\tilde{x}$ used by selfish agents such that each edge along that path is non-decreasing in flow from $x$, which we denote $p_{\rm{s}}$.

\begin{lemma}\label{lem:Path_Ordering}
    For all paths $p \in \mathcal{P}^{\rm{a}}$ and paths $q \in \mathcal{P}^{\rm{s}}$ such that $x_p^{\rm{a}} > 0$ and $x_q^{\rm{s}} > 0$, the following holds:
    \begin{align}
        x_p &\geq \tilde{x}_p,
        \\
        x_q &\leq \tilde{x}_q.
    \end{align}
\end{lemma}

Lemma~\ref{lem:Path_Ordering} demonstrates that the path flows for paths used by altruists in $x$ are guaranteed to be non-increasing in $\tilde{x}$, and the paths flows for paths used by selfish agents in $x$ are guaranteed to be non-decreasing in $\tilde{x}$.

\textit{Proof of Theorem~\ref{thm:Main}:}
Recall, $G$ and $\tilde{G}$ are identical games, except $\tilde{G}$ corresponds to changing $\varepsilon$ altruistic agents to selfish agents, where $0 < \varepsilon < x_p^{\rm{a}}$ for all $p \in \mathcal{P}^{\rm{a}}$ such that $x_p^{\rm{a}} > 0$.
That is, $r^{\rm{a}} - \varepsilon$ agents are altruistic and commute according to the marginal-cost of a path in $\tilde{G}$, and $r^{\rm{s}} + \varepsilon$ are selfish and commute according to the actual cost associated with a path in $\tilde{G}$.
Since $x$ is a Nash flow for altruists in $G$, it follows that for any $p \in \mathcal{P}^{\rm{a}}$ such that $x_p^{\rm{a}} > 0$ and any $q \in \mathcal{P}^{\rm{s}}$ such that $x_q^{\rm{s}} > 0$
\begin{equation}
    \ell_p^{\rm{mc}}(x) \leq \ell_q^{\rm{mc}}(x).
\end{equation}
Now, define
\begin{equation}
    \overline{p} \coloneqq \arg\min_{p \in \mathcal{P}^{\rm{s}}} \{\ell_p^{\rm{mc}}(x) : x_p^{\rm{s}} > 0\}.
\end{equation}
Then $\ell_p^{\rm{mc}}(x) \leq \ell_{\overline{p}}^{\rm{mc}}(x)$.
Next, Lemma~\ref{lem:Path_Ordering} implies that if a path flow increases from $x$ to $\tilde{x}$, then selfish agents use that path in $\tilde{x}$, and similarly, if a path flow decreases from $x$ to $\tilde{x}$, then altruistic agents use that path in $\tilde{x}$.
Thus, we are ready to bound the subgradient:
\begin{align}
    \frac{\partial}{\partial \varepsilon} \mathcal{L}(x(\varepsilon)) &= \nabla_x \mathcal{L}(x(\varepsilon)) \cdot \frac{\partial}{\partial \varepsilon}x(\varepsilon)
    \\ &\geq
    \varepsilon^* \leftbracket \ell_{\overline{p}}^{\rm{mc}}(x) - \ell_{p_{\rm{a}}}^{\rm{mc}}(x) \rightbracket
    \geq 0,
\end{align}
where $\varepsilon^* = \min \{ |\delta_{\rm{a}}|, |\delta_{\rm{s}}|\}$ for 
\begin{align}
    \delta_{\rm{a}} &= \min_{p \in \mathcal{P}}\{|\tilde{x}_p - x_p| : x_p^{\rm{a}} > 0\}
    \\
    \delta_{\rm{s}} &= \min_{p \in \mathcal{P}}\{|\tilde{x}_p - x_p| : x_p^{\rm{s}} > 0\}.
\end{align}
Since the subgradient at $x$ in the direction of $\tilde{x}$ is non-negative, it follows that $\mathcal{L}(x) \leq \mathcal{L}(\tilde{x})$, as desired.\hfill $\blacksquare$

\section{Networks that are not Braess-Resistant and Symmetric are Susceptible to Perversity}\label{sec:CounterExamples}

\begin{figure*}[t]
    \centering
        \begin{subfigure}[t]{.5\textwidth}
            \includegraphics[width=\textwidth]{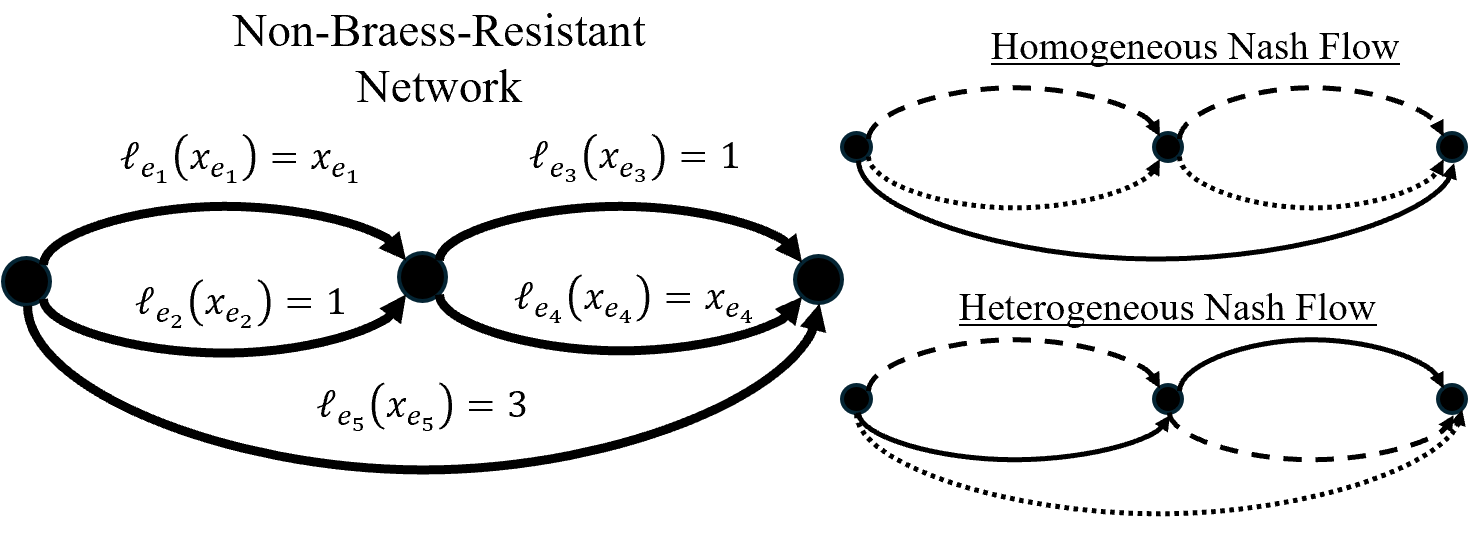}
            \caption{}
            \label{subfig:NBR}
        \end{subfigure}\hfill
        \begin{subfigure}[t]{.5\textwidth}
            \includegraphics[width=\textwidth]{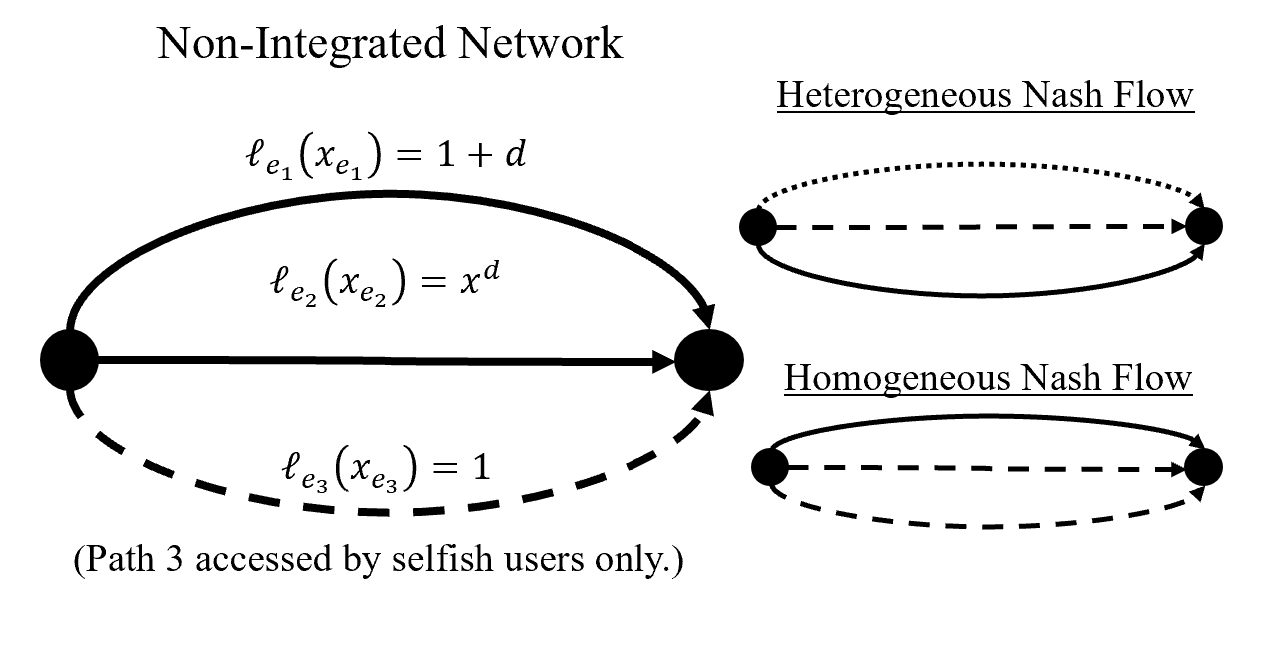}
            \caption{}
            \label{subfig:NI}
        \end{subfigure}\hfill
        \caption{Figure~\ref{subfig:NBR} is an example of the perversity that arises if a network is not Braess-resistant.
        The path set is $\mathcal{P} = \{(e_1, e_3), (e_1, e_4), (e_2, e_4), e_5\}$, so traffic travelling along $e_2$ cannot use $e_3$.
        In this example, two units of traffic are routed; when all traffic is selfish, half of the population routes along $(e_1, e_3)$, and the other half routes along $(e_2, e_4)$, for total latency $\mathcal{L}(\bar{x}) = 4$.
        Now, when half of traffic is altruistic, and the other half is selfish, the selfish agents will use $(e_1, e_4)$, and altruistic traffic will prefer $e_5$, resulting in total latency $\mathcal{L}(x) = 5$.
        Figure~\ref{subfig:NI} provides an example of the perversity that can arise if the path set is not symmetric.
        Again, two units of traffic traverse the network, and altruists have access to the top two edges while selfish agents have access to all three.
        If all traffic is selfish, the bottom two edges are used resulting in total latency $\mathcal{L}(\bar{x}) = 2$.
        Now, if half the population becomes altruistic, altruists now prefer the top edge, and the total latency becomes $\mathcal{L}(x) = 2 + d \geq 2 = \mathcal{L}(\bar{x})$.}
        \label{fig:Counterexamples}
        \vspace{-2mm}
\end{figure*}

Here, we present a short discussion on the network inefficiencies that arise if a network is not both Braess-resistant and symmetric.
It is well-known that networks which are not series-parallel can experience increased total congestion in the presence of altruism.
Hence, we maintain the series-parallel assumption, and provide networks that have the potential to have worsened total congestion in the presence of altruism.

\subsection{Networks which are not Braess-resistant}
Consider sending $2$ units of traffic across a series-parallel network consisting of five edges, so that the path set is $\mathcal{P} = \{(e_1,e_3), (e_1, e_4), (e_2,e_4), e_5\}$; notice that $(e_2, e_4)$ is not a path available to agents, hence $\mathcal{P}$ is not Braess-resistant.
If all traffic is selfish, then for Nash flow $\bar{x}$, $1$ unit of traffic uses path $(e_1, e_3)$, and the other half of traffic uses $(e_2, e_4)$, for a common latency $\Lambda_{\rm{s}}(\bar{x}) = 2$ (paths $(e_1, e_4)$ and $e_5$ have a latency of $4$ and $3$, respectively).
If half of traffic becomes altruistic, and the rest selfish (i.e. $r^{\rm{a}} = r^{\rm{s}} = 1$), then for a Nash flow $x$, selfish agents use path $(e_1, e_4)$ for a common latency of $\Lambda_{\rm{s}}(x) = 2$, and altruists use path $e_5$ for a common latency of $\Lambda_{\rm{a}}(x) = 3$.
Straightforward computation shows that $\mathcal{L}(\bar{x}) = 4$, and $\mathcal{L}(x) = 5$.
This example is shown in Figure~\ref{subfig:NBR}.

\subsection{Networks which are not Symmetric}
Consider sending $2$ units of traffic across a parallel network consisting of three edges, so that $\mathcal{P} = \{e_1, e_2, e_3\}$, $\mathcal{P}^{\rm{a}} = \{e_1, e_2\}$ and $\mathcal{P}^{\rm{s}} = \mathcal{P}$.
The latency for the paths are $\ell_{e_1}(x_{e_1}) = 1 + d$, $\ell_{e_2}(x_{e_2}) = x^d$, and  $\ell_{e_3}(x_{e_3}) = 1$, respectively. 
Now, a Nash flow for when all agents are selfish is $\bar{x} = \{0, 1, 1\}$.
This flow is obviously feasible, and Nash conditions are satisfied as
$\ell_{e_2}(1) = \ell_{e_3}(1) < \ell_{e_1}(0)$.
Now, if half of traffic becomes altruistic, and the rest selfish (i.e. $r^{\rm{a}} = r^{\rm{s}} = 1$), a resulting Nash flow is $x = \{r^{\rm{a}}, r^{\rm{s}}, 0\}$.
It is clearly feasible, and the cost altruistic traffic experiences is a Nash flow as
$\ell_{e_1}^{\rm{mc}}(r^{\rm{a}}) 
\leq \ell_{e_2}^{\rm{mc}}(r^{\rm{s}})$.
Finally, straightforward computation shows that $\mathcal{L}(\bar{x}) = 2$, and $\mathcal{L}(x) = 2 + d$.
This example is shown in Figure~\ref{subfig:NI}.

\section{CONCLUSIONS}\label{sec:Conclusion}

We have provided a specific set of conditions for network topology (in addition to series-parallel) and network access among populations so that an increase in the fraction of altruistic traffic relative to the total population is guaranteed to lower total congestion.
Future work will focus on extending to heterogeneous populations that consist of more than only strictly selfish and completely altruistic sub-populations, so the effects of multiple partially altruistic sub-populations on total congestion can be better understood.

\section*{APPENDIX}

\begin{figure*}[t]
    \centering
        \begin{subfigure}[t]{.33\textwidth}
            \includegraphics[width=\textwidth]{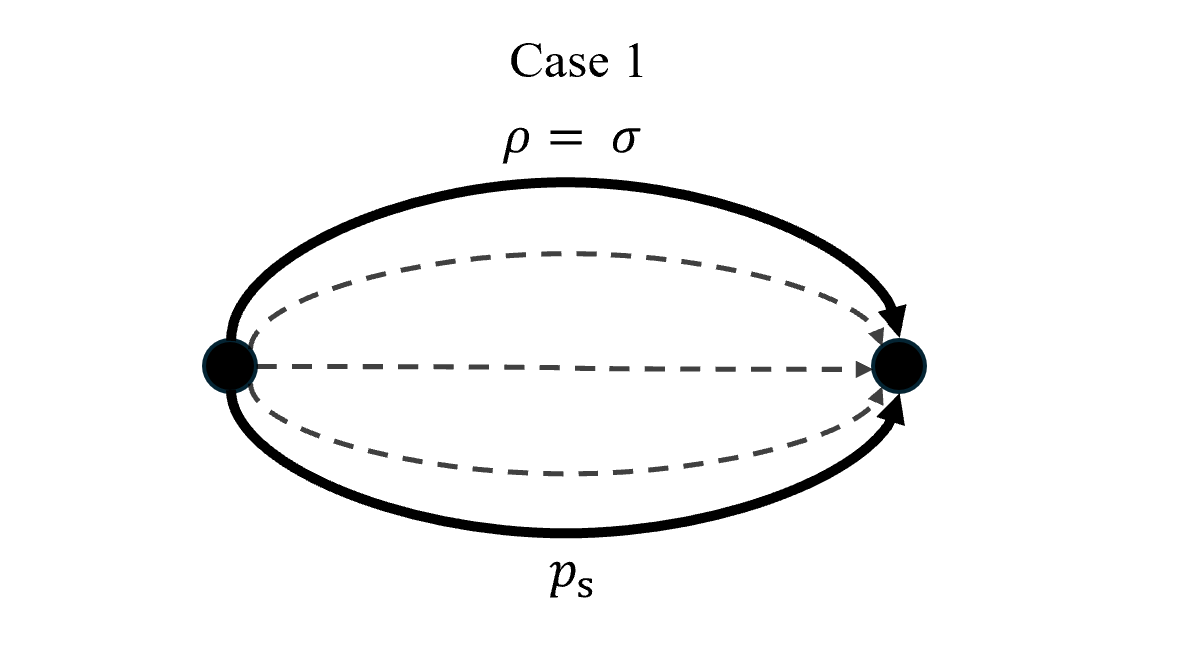}
            \caption{}
            \label{subfig:Case_1}
        \end{subfigure}\hfill
        \begin{subfigure}[t]{.33\textwidth}
            \includegraphics[width=\textwidth]{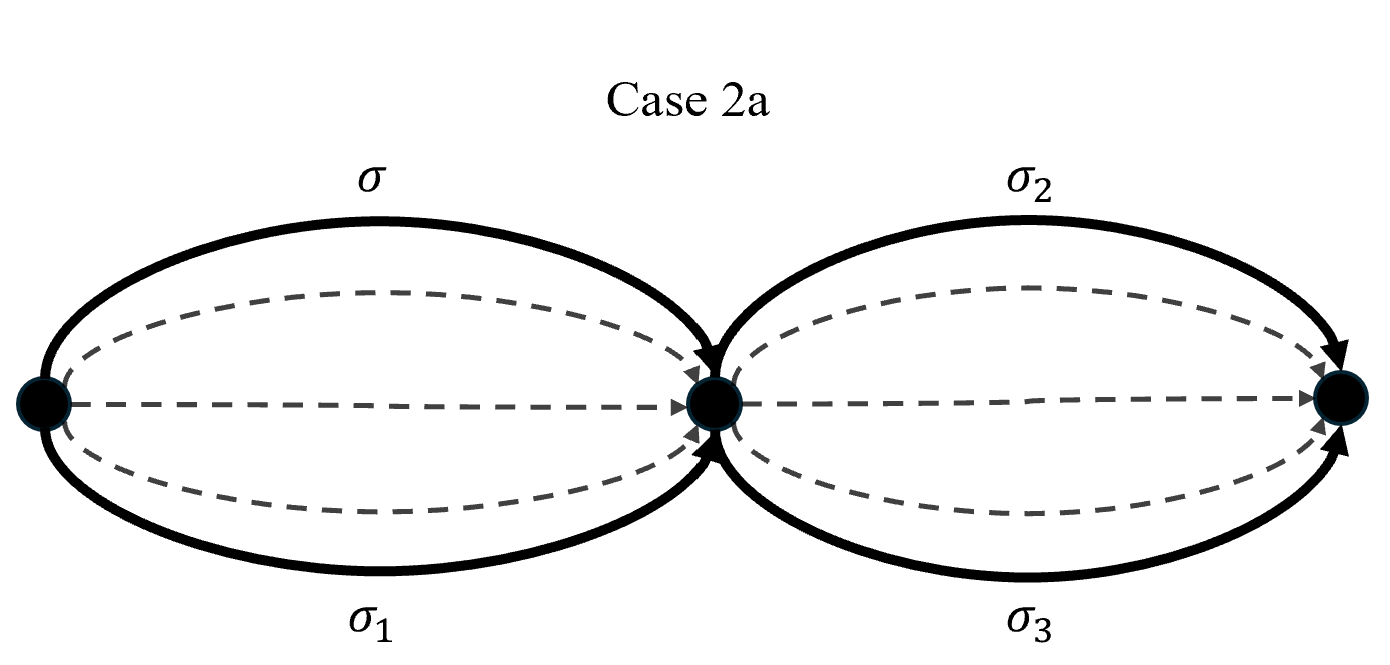}
            \caption{}
            \label{subfig:Case_2}
        \end{subfigure}\hfill
        \begin{subfigure}[t]{.33\textwidth}
            \includegraphics[width=\textwidth]{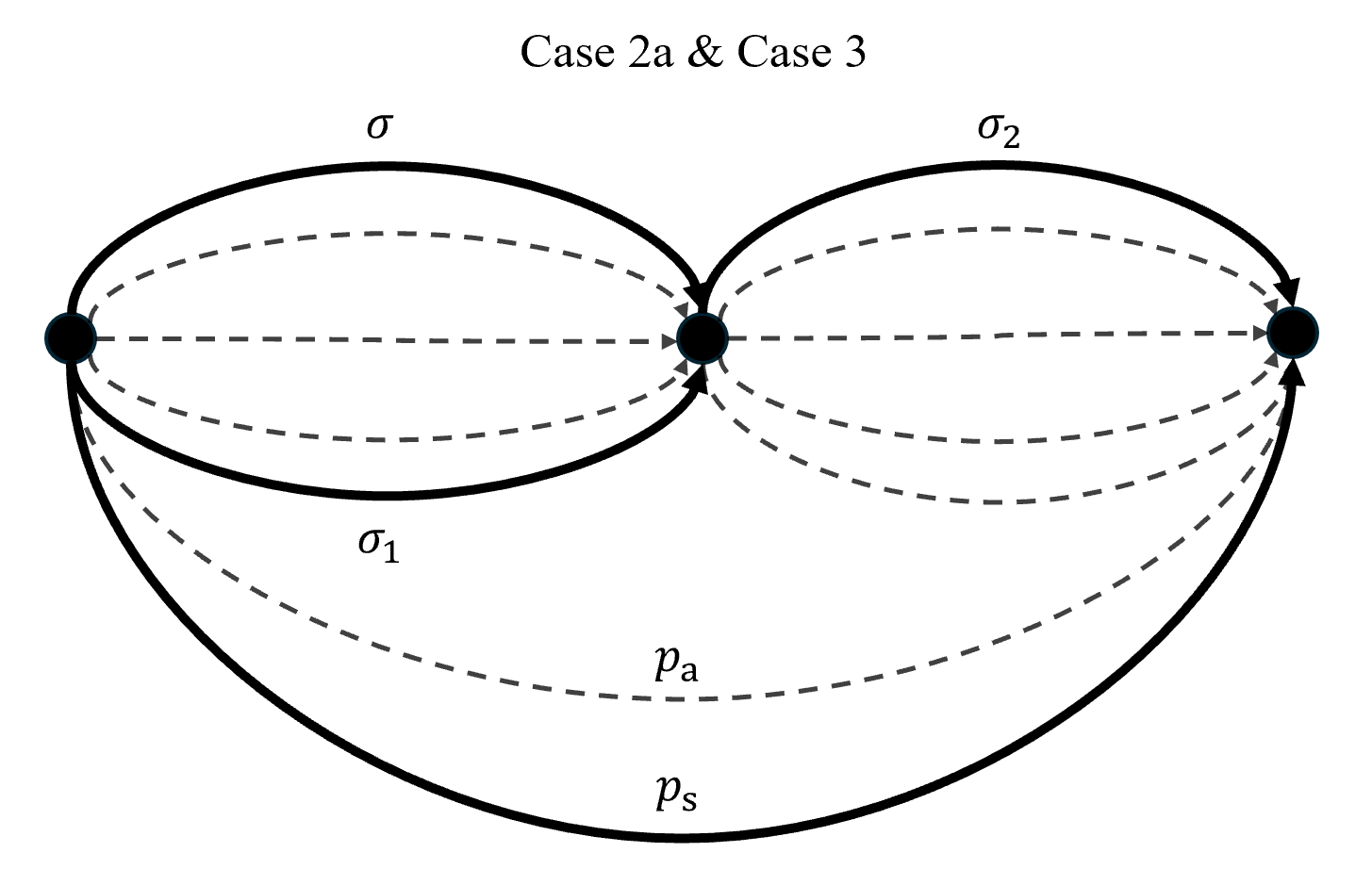}
            \caption{}
            \label{subfig:Case_3}
        \end{subfigure}
        \caption{Geometric interpretation for the proof of Lemma~\ref{lem:Path_Ordering}.
        Figure~\ref{subfig:Case_1} represents the network that arises in \textit{Case 1}.
        Here, there exists a path, used by selfish agents in $x$, $\rho$, that has every edge strictly decreasing from $x$ to $\tilde{x}$, and so the cost of the path goes down.
        However, we know from Corollary~\ref{cor:p_a_AND_p_s} that there is a min-cost path, $p_{\rm{s}}$, in $\tilde{x}$, such that each edge in the path is non-decreasing, so the cost of the path is non-decreasing. 
        This implies that the cost of $p_{\rm{s}}$ is strictly greater that the cost of $\rho$, contradicting that $p_{\rm{s}}$ is a min-cost path.
        Figure~\ref{subfig:Case_2} presents a similar argument.
        Here, selfish agents using $p_{\rm{s}} = \sigma_1 + \sigma_3$ are better off using $\sigma + \sigma_3$ or $\sigma + \sigma_2$, both of which are possible since the path set is Braess-resistant.
        Finally, figure~\ref{subfig:Case_3} shows the [sub]-network that arises if $p_{\rm{s}}$ shares no common vertices with $\rho$.
        Because the flow is strictly decreasing on $\sigma$, and non-decreasing on $\sigma_2$, there must exist a sub-path $\sigma_1$ that shares incident vertices with $\sigma$.
        It can be deduced that altruists are using $\sigma_1 + \sigma_2$, and that their cost in $\tilde{x}$ has either strictly gone up from $x$, or that the path $p^\prime = \sigma_1 + \sigma_2$ and $p_{\rm{a}}$ are constant.
        Either case produces a contradiction.}
        \label{fig: path_proof}
        \vspace{-2mm}
\end{figure*}

\textit{Proof of Lemma~\ref{lem:Milchtaich_Extension}:}
We slightly abuse notation, and refer to the network for a game $G$ by $G$ itself.
The proof proceeds by induction on the number of edges.
For a network with a single edge, the result is trivial.
Next, consider a series-parallel network with two or more edges.
The induction hypothesis asserts that the lemma holds for any two flow vectors in every series-parallel network with a smaller number of edges than $G$.
It is known that $G$ can be constructed from connecting two series-parallel networks, $G^\prime$, and $G^{\prime\prime}$, in series or in parallel.
We first consider $G^\prime$, and $G^{\prime\prime}$ are connected in series, so that the destination node of $G^\prime$ is the source node of $G^{\prime\prime}$.
The path sets of $G$ and $G^\prime$, $\mathcal{P}$ and $\mathcal{P}^\prime$, are connected by $\mathcal{P}^\prime = \{p_{ot^\prime} | p \in \mathcal{P}\}$, where $t^\prime$ is the destination node of $G^\prime$.
Thus, the paths of $\mathcal{P}^\prime$ are sub-paths of $\mathcal{P}$.
Now, every flow vector $f$ for $G$ induces a flow vector $f^\prime$ for $G^\prime$, which is given by $f^\prime = (f_{p^\prime})_{p^\prime \in \mathcal{P}^\prime}$.
By definition, the flow $f_e^\prime$ on each edge $e$ in $G^\prime$ is the total flow on all the paths in $G^\prime$ containing $e$, that is
\begin{equation}
    f_e^\prime = \sum_{p^\prime \in \mathcal{P}^\prime, p^\prime \ni e} f_{p^\prime}.
\end{equation}
Notice this simply implies $f_e^\prime = f_e$, the flow on edge $e$ in $f$.
Thus, if $\hat{f}$ and $\tilde{f}$ are flow vectors for $G$ satisfying $||\hat{f}||_1 \geq ||\tilde{f}||_1$, $||\hat{f}^\theta||_1 > 0$, then $||\hat{f}^\prime||_1 \geq ||\tilde{f}^\prime||$, $||{f^\theta}^\prime||_1 > 0$ also holds for the induced flow vectors $\hat{f}^\prime$ and $\tilde{f}^\prime$ for $G^\prime$.
Hence, by the induction hypothesis, there is a path $p^\prime$ in $\mathcal{P}^\prime$ such that, for all $e \in p^\prime$, \mbox{$\hat{f}_e^\theta \geq \tilde{f}_e^\theta$}, and $\hat{f}_e^\theta > 0$.
By a similar argument, there exists a path $p^{\prime\prime}$ in $\mathcal{P}^{\prime\prime}$ such that identical inequalities hold.
Since $\mathcal{P}$ is Braess resistant, there is a path $p \in \mathcal{P}$ such that $p = p^\prime + p^{\prime\prime}$ that satisfies the same set of inequalities.
This completes the proof in the case that $G$ is the result of connecting two series-parallel networks in series.
Next consider $G$ is the result of connecting $G^\prime$ and $G^{\prime\prime}$ in parallel, so that $o$ and $t$ are also the source and destination nodes of $G^\prime$ and $G^{\prime\prime}$.
It follows by definition that the path sets of $G^\prime$ and $G^{\prime\prime}$, $\mathcal{P}^\prime$ and $\mathcal{P}^{\prime\prime}$, are disjoint, and $\mathcal{P} = \mathcal{P}^\prime \cup\mathcal{P}^{\prime\prime}$.
Any feasible flow $f$ of $G$ induces flow vectors $f^\prime$ and $f^{\prime\prime}$ for $G^\prime$ and $G^{\prime\prime}$, which are given by $f^\prime = (f_{p^\prime})_{p^\prime \in \mathcal{P}^\prime}$ and $f^{\prime\prime} = (f_{p^{\prime\prime}})_{p^{\prime\prime} \in \mathcal{P}^{\prime\prime}}$.
Hence, the induced flow vectors give the flow $f_e^\prime$ on either each edge $e$ in $G^\prime$, or $f_e^{\prime\prime}$ on each edge $e$ in $G^{\prime\prime}$, respectively.
That is,
\begin{equation}
    f_e^\prime = \sum_{p^\prime \in \mathcal{P}^\prime, p^\prime \ni e} f_{p^\prime},
\end{equation}
or similarly for $G^{\prime\prime}$:
\begin{equation}
    f_e^{\prime\prime} = \sum_{p^{\prime\prime} \in \mathcal{P}^{\prime\prime}, p^{\prime\prime} \ni e} f_{p^{\prime\prime}}.
\end{equation}
Then, either $f_e^\prime = f_e$ or $f_e^{\prime\prime} = f_e$.
Now, $||f^\prime||_1 + ||f^{\prime\prime}||_1 = \sum_{p^\prime \in \mathcal{P}^\prime} f_{p^\prime} + \sum_{p^{\prime\prime} \in \mathcal{P}^{\prime\prime}} f_{p^{\prime\prime}} = ||f||_1$.
It follows from this equation that if $||\hat{f}||_1 \geq ||\tilde{f}||_1$ and $||\hat{f}^\theta||_1 > 0$, then $||\hat{f}^\prime||_1 \geq ||\tilde{f}^\prime||_1$ and $||\hat{f}^{\theta^\prime}||_1 > 0$, or $||f^{\prime\prime}|| \geq ||\tilde{f}^{\prime\prime}||_1$ and $||\hat{f}^{\theta^{\prime\prime}}||_1 > 0$ hold.
Thus, by the induction hypothesis, there is some path in $\mathcal{P}^\prime$ or $\mathcal{P}^{\prime\prime}$ (and hence, in $\mathcal{P}$) such that $\hat{f}_e^\theta \geq \tilde{f}_e^\theta$ and $\hat{f}_e^\theta > 0$ for each $e \in p$.\hfill $\blacksquare$

\textit{Proof of Corollary~\ref{cor:p_a_AND_p_s}:}
The existence of $p_{\rm{a}}$ and $p_{\rm{s}}$ follows immediately from Lemma~\ref{lem:Milchtaich_Extension}.
By~\cite[Lemma 3]{milchtaichNetworkTopologyEfficiency2006}, $p_{\rm{a}}$ is a minimum cost path for altruists in $x$, and $p_{\rm{s}}$ is a minimum cost path for selfish agents in $\tilde{x}$, so we have the following for $p_{\rm{a}}$:
\begin{align}
    \Lambda_{\rm{a}}(x) &= \ell_{p_{\rm{a}}}^{mc}(x) 
    \\ &= 
    \sum_{e \in {p_{\rm{a}}}} \ell_e^{mc}(x_e) 
    \\ &\geq
    \sum_{e \in p} \ell_e^{mc}(\tilde{x}_e) 
    \\ &=
    \ell_{p_{\rm{a}}}^{mc}(\tilde{x})
    \\ &\geq
    \Lambda_{\rm{a}}(\tilde{x})
\end{align}
and the following for $p_{\rm{s}}$:
\begin{align}
    \Lambda_{\rm{s}}(x) &\leq \ell_{p_{\rm{s}}}(x) 
    \\ &= \sum_{e \in {p_{\rm{s}}}} \ell_e(x_e) 
    \\ &\leq \sum_{e \in {p_{\rm{s}}}} \ell_e(\tilde{x}_e) 
    \\ &=
    \ell_{p_{\rm{s}}}(\tilde{x})
    \\ &=
    \Lambda_{\rm{s}}(\tilde{x}).
\end{align}
Hence, we have $\Lambda_{\rm{a}}(x) \geq \Lambda_{\rm{a}}(\tilde{x})$ and $\Lambda_{\rm{s}}(x) \leq \Lambda_{\rm{s}}(\tilde{x})$, as desired.\hfill $\blacksquare$

\textit{Proof of Lemma~\ref{lem:Path_Ordering}:} 
Suppose by way of contradiction that the claim is false.
Then there exists a path $\rho \in \mathcal{P}^{\rm{s}}$ such that $x_\rho^{\rm{s}} > 0$ and $x_\rho > \tilde{x}_\rho$.
Without loss of generality, there exists a sub-path $\sigma \in \rho$, such that $\sigma$ has one or more edges, where $x_e > \tilde{x}_e$ for all $e \in \sigma$.
We may also assume that $\sigma$ is the first such sub-path of $\rho$ that satisfies the above condition.
\\ \textit{Case 1:} We first consider the case that $\sigma = \rho$ (i.e. $x_e > \tilde{x}_e$ for all $e \in \rho$).
Then $\ell_\rho(x) = \ell_\sigma(x) > \ell_\sigma(\tilde{x}) = \ell_\rho(\tilde{x})$.
Now, recall that $p_{\rm{s}}$ is a min-cost path for selfish agents in $\tilde{x}$ and each edge in $p_{\rm{s}}$ is non-decreasing.
Hence $\ell_{p_{\rm{s}}}(x) \leq \ell_{p_{\rm{s}}}(\tilde{x})$.
But then we have the following:
\begin{align}
    \ell_\rho(x) &\leq \ell_{p_{\rm{s}}}(x) 
    \\ &\leq
    \ell_{p_{\rm{s}}}(\tilde{x}) 
    \\ &\leq
    \ell_\rho(\tilde{x}) 
    \\ &< 
    \ell_\rho(x).
\end{align}
Which is a contradiction.
\\ \textit{Case 2:} We next consider the case that $\rho$ consists of two sub-paths.
Then $\rho = \sigma + \sigma_2$, where $x_e \leq \tilde{x}_e$ for all $e \in \sigma_2$.
First, consider the case that an edge in $p_{\rm{s}}$ is incident to the vertex connecting $\sigma$ with $\sigma_2$.
Then $p_{\rm{s}} = \sigma_1 + \sigma_3$, where $\sigma$ shares incident vertices with $\sigma_1$, and $\sigma_2$ shares incident vertices with (and is possibly equal to) $\sigma_3$.
Then, since $\mathcal{P}$ is Braess-resistant, there exists a path $p^\prime \in \mathcal{P}$ such that $p^\prime = \sigma + \sigma_3$.
Then 
\begin{align}
    \ell_{p^\prime}(\tilde{x}) &= \ell_{\sigma}(\tilde{x}) + \ell_{\sigma_3}(\tilde{x}) 
    \\ &<
    \ell_{\sigma_1}(\tilde{x}) + \ell_{\sigma_3}(\tilde{x}) 
    \\ &=
    \ell_{p_s}(\tilde{x}).
\end{align}
Since $\mathcal{P}^{\rm{s}}$ is symmetric, $\tilde{x}$ is not a Nash flow for selfish agents, a contradiction.
Next, consider that no edge in $p_{\rm{s}}$ is incident to any vertex in $\rho$, except $o$ and $t$.
Now, because the flow on $\sigma_2$ is non-decreasing, and the flow on $\sigma$ is strictly decreasing, there exists a decomposition of the network containing incident sub-paths with $\sigma$ that has increased total flow from $x$ to $\tilde{x}$ (otherwise, $\sigma_2$ does not have non-decreasing flow).
Thus, there is a path $p^\prime = \sigma_1 + \sigma_2$, such that $x_e < \tilde{x}_e$ for all $e \in \sigma_1$ (this fact also follows from Lemma~\ref{lem:Milchtaich_Extension}).
Hence $p^\prime$ is a min-cost path for either altruists or selfish agents.
Notice that $p^\prime$ is a min-cost path for altruists ($p^\prime$ being a min-cost path for altruists would imply $\tilde{x}$ is not a Nash flow, since selfish agents could unilaterally switch to $\sigma$ from $\sigma_1$, and have strictly improved their cost).
Hence 
\begin{align}
    \ell^{mc}_{p^\prime}(x) &= \ell^{mc}_{\sigma_1}(x) + \ell^{mc}_{\sigma_2}(x) 
    \\ &\leq \label{eq:induction}
    \ell^{mc}_{\sigma_1}(\tilde{x}) + \ell^{mc}_{\sigma_2}(\tilde{x}) 
    \\ &= 
    \ell^{mc}_{p^\prime}(\tilde{x}).
\end{align}
But then, either $\ell^{mc}_{p_{\rm{a}}}(\tilde{x}) \leq \ell^{mc}_{p_{\rm{a}}}(x) \leq \ell^{mc}_{p^\prime}(x) < \ell^{mc}_{p^\prime}(\tilde{x})$, contradicting that $p^\prime$ is used by altruists in $\tilde{x}$, or $\ell^{mc}_{p_{\rm{a}}}$ and $\ell^{mc}_{p^\prime}$ are constant and equal to one another, in which case all altruists are using $p^\prime$ arbitrarily, and can be rerouted to $p_{\rm{a}}$, but then $x_e > \tilde{x}_e$ for all $e \in \rho$, which leads to the same contradiction as \textit{Case 1}.
\\ \textit{Case 3:} Lastly, we consider the case that $\rho$ is a series of sub-paths that are constructing so that each sub-path in $\rho$ is a series of strictly decreasing edges, or non-decreasing edges.
That is
\begin{equation}
    \rho = \sum_{i: \rho \ni \sigma_i} \sigma_i,
\end{equation}
where we may assume without loss of generality that \mbox{$x_e > \tilde{x}_e$} for each $\sigma_{2i}$ and $x_e \leq \tilde{x}_e$ for each $\sigma_{2i + 1}$ (the edge flows are strictly decreasing on even sub-paths, and are non-decreasing on odd sub-paths in $\rho$).
Now, by inductively applying the same reasoning from \textit{Case 2}, part b), selfish agents must be sharing each odd sub-path in $\rho$ with altruists, and so there must exist a sub-path (call it $\sigma_{2i}^\prime$) sharing incident vertices with each even sub-path in $\rho$ such that $x_e < \tilde{x}_e$ and $\tilde{x}_e^{\rm{a}} > 0$ for all $e \in \sigma_{2i}^\prime$.
Thus, since $\mathcal{P}$ is Braess-resistant, there exists a path
\begin{equation}
    p^\prime = \sum_{i: \rho \ni 2i+1}\sigma_i + \sum \sigma_{2i}^\prime,
\end{equation}
where each $x_e \leq \tilde{x}_e$ and $\tilde{x}_e^{\rm{a}} > 0$ for each $e \in p^\prime$.
Thus $p^\prime$ is a min-cost path for altruists in $\tilde{x}$, and so $\ell_{p^\prime}^{\rm{mc}}(x) \leq \ell_{p^\prime}^{\rm{mc}}(\tilde{x})$.
Hence, in a similar fashion as \textit{Case 2}, part b), either $\tilde{x}$ is not a Nash flow, or $\ell^{mc}_{p_{\rm{a}}}$ and $\ell^{mc}_{p^\prime}$ are constant and equal to one another, both of which lead to contradictions.
Figure~\ref{fig: path_proof} presents a geometric interpretation of the proof.

Similar arguments show that $x_p \geq \tilde{x}_p$ for all $p \in \mathcal{P}^{\rm{a}}$ such that $x_p^{\rm{a}} > 0$.
\hfill $\blacksquare$



\end{document}